\documentclass[article,twocolumn,showpacs,aps,prb,floatfix]{revtex4-1}
\usepackage{graphicx,subfigure,epsfig,verbatim,psfrag,amsmath,amssymb,color}
\usepackage{indentfirst}
\usepackage{textcomp}
\usepackage{latexsym}
\usepackage{epstopdf}
\usepackage{amssymb}
\usepackage{amsmath}
\usepackage{lipsum}
\usepackage{soul}
\usepackage{xcolor}
\makeatletter
\begin{document}
\title{An exactly solvable BCS-Hubbard Model in arbitrary dimensions}
\author{Zewei Chen}
\email{zchenal@connect.ust.hk}
\author{Xiaohui Li}
\author{Tai Kai Ng}
\affiliation{Department of Physics, Hong Kong University of Science and Technology, Clear Water Bay, Hong Kong, China}
\begin{abstract}
We introduce in this paper an exact solvable BCS-Hubbard model in arbitrary dimensions. The model describes a $p$-wave BCS superconductor with equal spin pairing moving on a bipartite (cubic, square etc.) lattice with on site Hubbard interaction $U$. We show that the model becomes exactly solvable for arbitrary $U$ when the BCS pairing amplitude $\Delta$ equals the hopping amplitude $t$. The nature of the solution is described in detail in this paper. The construction of the exact solution is parallel to the exactly solvable Kitaev honeycomb model for $S=1/2$ quantum spins and can be viewed as a generalization of Kitaev's construction to $S=1/2$ interacting lattice fermions. The BCS-Hubbard model discussed in this paper is just an example of a large class of exactly solvable lattice fermion models that can be constructed similarly.
\end{abstract}
\maketitle
\emph{Introduction}---Exact solutions of quantum interacting-particle models in dimensions $>1$ are rare and are important resources for understanding the physics of strongly-correlated systems in dimensions $>1$\cite{march2016exactly,Mahan2000,RevModPhys.76.643,richardson1964exact,richardson1965exact,richardson1968exactly,sierra2000conformal,schechter2001thermodynamic,PhysRevLett.90.016803,vijay2017generalization,PhysRevLett.102.217202,PhysRevLett.66.1383,ghosh1998exactly}. More recently, a major advance in understanding the mathematics of topological order was put forth by the introduction of the exactly solvable Toric Code\cite{toric} and honeycomb\cite{kitaev2006anyons,PhysRevLett.98.247201,chen2008exact,feng2007topological} models by Kitaev  and their generalizations\cite{PhysRevLett.99.247203,PhysRevB.83.180412,1742-5468-2014-10-P10022,PhysRevB.81.125134,PhysRevB.85.155119,PhysRevB.79.134427,PhysRevB.76.180404,kells2011kaleidoscope,PhysRevLett.114.157202}. The exact solvability of the Kitaev honeycomb model is a result of the existence of an infinite number of conserved quantities (for an infinite lattice) in the model. In this paper, we show that Kitaev's construction can be generalized to a class of $S=1/2$ lattice fermion models that describe p-wave BCS superconductors with equal spin pairing and with on site Hubbard interaction $U$. The generalization is based on the observation that the Kitaev honeycomb lattice model can be expressed in terms of spinless fermion model\cite{chen2008exact,feng2007topological} {\em via} a Jordan-Wigner transformation. 
The ``generalized" lattice fermion models carry both ``quasi-particle" and ``solitonic" excitations as in the Kitaev honeycomb model except that the solitonic excitations are in general non-topological in lattice fermion models.

 To illustrate, we consider in this paper a particular BCS-Hubbard model with equal spin pairing \cite{lee1997anisotropy,PhysRevLett.52.679,maeno1994superconductivity} on cubic (3D) and square (2D) lattices. The more general constructions are discussed at the end of the paper.

\begin{figure}
\centering
\includegraphics[width=0.5\columnwidth]{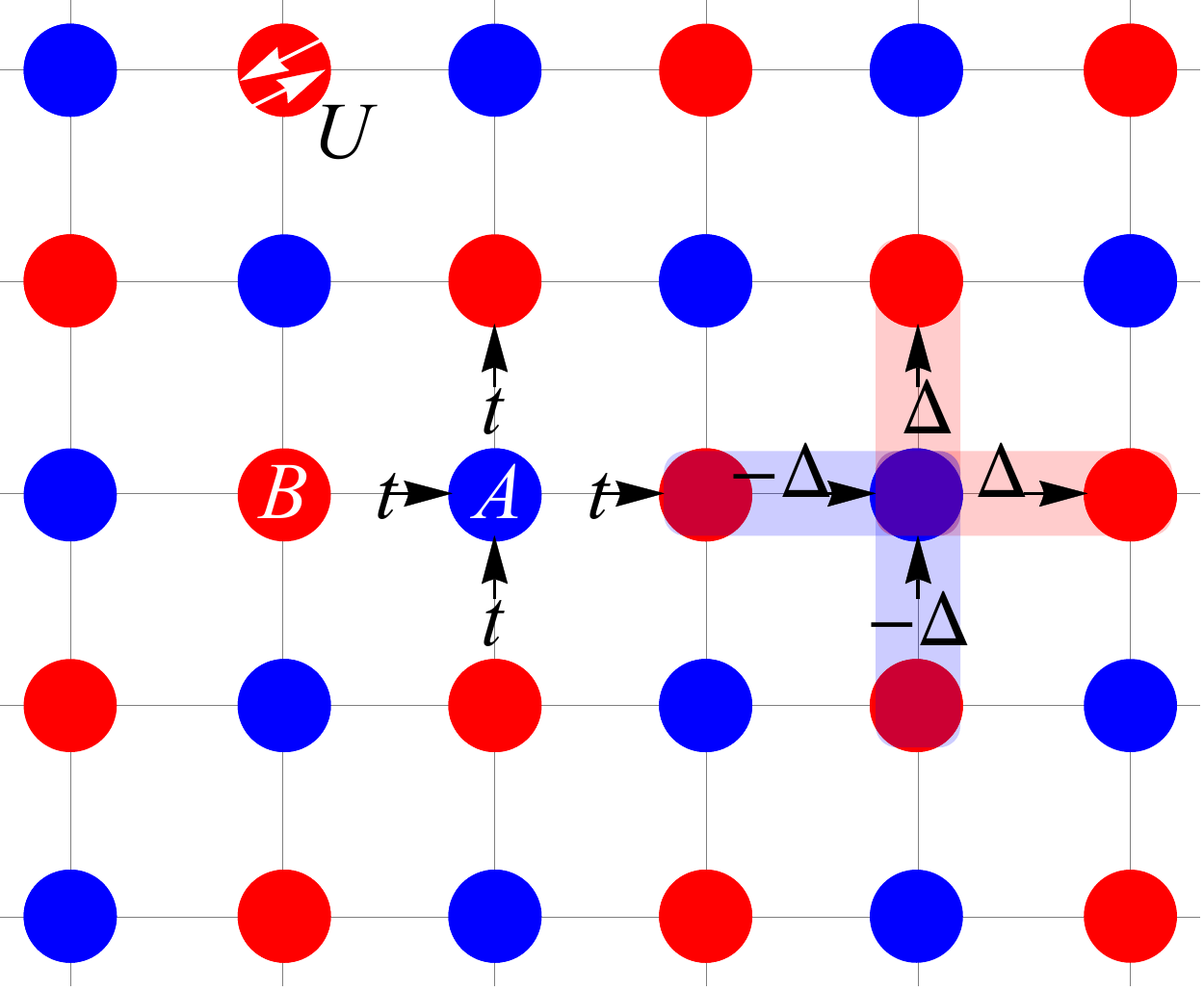}
\caption{BCS-Hubbard model on square lattice, naming the blue dot as $A$ sub-lattice and red dot $B$ sub-lattice. The hopping potential $t$ is uniform along all the nearest neighbor bond, that is $t_{A\rightarrow B} = t_{B\rightarrow A}$, while the ESP potential has a staggered form which explicitly is $\Delta_{A\rightarrow B} = -\Delta_{B\rightarrow A}$. $U$ represents the Hubbard onsite interaction.}\label{Fig:ModelPlot}
\end{figure}

\emph{Model}---The Hamiltonian for our BCS-Hubbard model is given by $H=H_0+H_{int}$, where
\begin{eqnarray}
\label{h}
H_0 & = & \sum_{\langle i,j\rangle,\sigma}\left(t_{ij}c^{\dagger}_{i\sigma}c_{j\sigma}+h.c. +\Delta_{ij} c^{\dagger}_{i\sigma}c^{\dagger}_{j\sigma}+h.c.\right) \\ \nonumber
H_{int} & = & U\sum_{l}(n_{l\uparrow}-{1\over2})(n_{l\downarrow}-{1\over2})
\end{eqnarray}
 where $\langle i,j\rangle$ describes nearest neighbor sites with $i\in A, j\in B$ being lattices sites belonging to different sublattices of the cubic or square lattice. $t_{ij}=t_{ji}$ and $\Delta_{ij}=-\Delta_{ji}$ are hopping matrix and BCS-pairing term between sites $i$ and $j$, respectively. The last term describes on-site Hubbard interaction $U$ where $l\in A,B$, i.e. all lattices sites where $n_{l\sigma}=c^{\dagger}_{l\sigma}c_{l\sigma}$. Notice that the BCS-pairing term describes equal spin pairing. We shall consider a pairing term $\Delta_{ij}$ which is positive when $i\in A, j\in B$, corresponding to a staggering nearest neighbor pairing field on the cubic and square lattices (see Fig.(\ref{Fig:ModelPlot}) ).
 The Hamiltonian \ (\ref{h}) is in general not solvable. In the following we shall show that it becomes exactly solvable when $\Delta_{ij}=t_{ij}=t$ where $t$ is a real number.

 \emph{Construction of exact solution}---To see how the model becomes exactly solvable we introduce Majorana fermion representation
\begin{eqnarray}
\label{majorana}
c_{i\sigma}=\eta_{i\sigma}+i\beta_{i\sigma}, c^{\dagger}_{i\sigma}=\eta_{i\sigma}-i\beta_{i\sigma}  \\ \nonumber
c_{j\sigma}=\beta_{j\sigma}+i\eta_{j\sigma}, c^{\dagger}_{j\sigma}=\beta_{j\sigma}-i\eta_{j\sigma}  \\ \nonumber
\end{eqnarray}
for fermions on $i\in A$ and $j\in B$ sublattices, respectively. It is straightforward to show that the Hamiltonian $H$ can be represented in terms of Majorana fermions $\eta$'s and $\beta$'s where
\begin{eqnarray}
\label{hma}
H_0 & \rightarrow & 4i\tilde{t}\sum_{\langle i,j\rangle,\sigma}\left(-\beta_{i\sigma}\beta_{j\sigma}+\delta\eta_{i\sigma}\eta_{j\sigma}\right) \\ \nonumber
H_{int} & \rightarrow & U\sum_{l}(2i\eta_{l\uparrow}\beta_{l\uparrow})(2i\eta_{l\downarrow}\beta_{l\downarrow})
\end{eqnarray}
where $t=\tilde{t}(1+\delta)$ and $\Delta=\tilde{t}(1-\delta)$. We notice that in the limit $t=\Delta$ (or $\delta=0$), the kinetic ($H_0$) term is expressed in terms of Majorana fermions $\beta$'s only and the $\eta$ terms are absent in $H_0$. In this limit,
\[
 {d\over dt}(\eta_{l\uparrow}\eta_{l\downarrow})={1\over i\hbar}[\eta_{l\uparrow}\eta_{l\downarrow}, H]=0,  \]
$\forall l$ and $(2i\eta_{l\uparrow}\eta_{l\downarrow})=D_l$ are constants of motion. Using the identities $(\eta_{l\sigma})^2=(\beta_{l\sigma})^2={1\over4}$, we obtain $(D_l)^2={1\over4}$ and $D_l=\pm{1\over2}$.

 Substituting $D_l$ into Eq.\ (\ref{hma}), we obtain in the limit $\delta\rightarrow0$,
\begin{equation}
\label{hma2}
H\rightarrow -4it  \sum_{\langle i,j\rangle,\sigma}\left(\beta_{i\sigma}\beta_{j\sigma}\right)-\sum_{l}(UD_l)(2i\beta_{l\uparrow}\beta_{l\downarrow})
\end{equation}
where $D_l$ are $C$-numbers. The Hamiltonian\ (\ref{hma2}) is quadratic and is exactly diagonalizable. The many-body eigenstates of the Hamiltonian are divided into different {\em solitonic} sectors characterized by different sets of  eigenvalues $\{D_l\}$. The ground state of the system is given by the set of $\{D_l\}$ with lowest energy. The construction of the exact solution is parallel to the construction of the exact solution of the spin-$1/2$ Kitaev honeycomb model\cite{kitaev2006anyons} when the model is expressed in terms of spinless fermions\cite{chen2008exact,feng2007topological}. We show here that the Kitaev construction can be extended to $S=1/2$ lattice fermions with Hubbard-type interaction rather straightforwardly.


 \emph{Properties of the exact solution (1) - U=0 limit}---We first study the solution of Hamiltonian\ (\ref{hma}) in the limit $U=0$. In this limit $H\rightarrow H_0$ describes a p-wave, ESP BCS superconductor with staggered nearest neighbor pairing fields and with chemical potential $\mu=0$, i.e. half-filled bands.

 It is convenient to ``re-fermionize" the Majorana fermions by introducing the composite fermions
 \begin{eqnarray}
\label{cfermion}
d_{l2} & = & \eta_{l\uparrow}+i\xi_l\eta_{l\downarrow}, d^{\dagger}_{l2}=\eta_{l\uparrow}-i\xi_l\eta_{l\downarrow}  \\ \nonumber
d_{l1} & = & \beta_{l\uparrow}-i\xi_l\beta_{l\downarrow}, d^{\dagger}_{l1}=\beta_{l\uparrow}+i\xi_l\beta_{l\downarrow}.  \\ \nonumber
\end{eqnarray}
where $\xi_l=+(-)1$ for $l\in A(B)-$ sublattices.

The transformation\ (\ref{majorana}) and\ (\ref{cfermion}) and be understood by introducing the fermions
\begin{eqnarray}
\label{fermiony}
c_{l\rightleftarrows} & = & {1\over\sqrt{2}}(c_{l\uparrow}\pm ic_{l\downarrow})
\end{eqnarray}
which represents fermions with spin pointing in $+(-)\hat{y}-$directions, respectively. The $d$ fermions are related to $c_{\rightleftarrows}$ by
\begin{subequations}
\label{bdgy}
\begin{eqnarray}
\label{bdgy1}
d_{l\rightarrow} & = & {1\over\sqrt{2}}(c_{l\rightarrow}+c^{\dagger}_{l\leftarrow})  \\ \nonumber
d_{l\leftarrow} & = & {1\over\sqrt{2}i}(c_{l\leftarrow}-c^{\dagger}_{l\rightarrow}),
\end{eqnarray}
and
\begin{eqnarray}
\label{bdgy2}
d_{l2}^{(A)}=d_{l1}^{(B)} & = & d_{l\rightarrow}  \\ \nonumber
d_{l1}^{(A)}=d_{l2}^{(B)} & = & d_{l\leftarrow},
\end{eqnarray}
\end{subequations}
where $d^{(A)}$ and $d^{(B)}$ are fermions on $A(B)$- sublattices, respectively. Eq.\ (\ref{bdgy}) represents a Bogoliubov-de Gennes transformation between the  $d$ and $c_{\rightleftarrows}$ fermions.

  As will be seen below and in next section, the $d$-fermions will form the {\em quasi-particles} for our model Hamiltonians. It's interesting to note from Eq.\ (\ref{bdgy1}) that
  \begin{subequations}
  \begin{equation}
  \label{c&s1}
  \langle c^{\dagger}_{l\leftarrow}c_{l\leftarrow}\rangle+\langle c^{\dagger}_{l\rightarrow}c_{l\rightarrow}\rangle = 1+i\langle (d^+_{l\rightarrow}d^{\dagger}_{l\leftarrow}-d_{l\leftarrow}d_{l\rightarrow})\rangle, 
  \end{equation}
  and
  \begin{equation}
  \label{c^s2}
  \langle c^{\dagger}_{l\leftarrow}c_{l\leftarrow}\rangle-\langle c^{\dagger}_{l\rightarrow}c_{l\rightarrow}\rangle =
  \langle d^{\dagger}_{l\leftarrow}d_{l\leftarrow}\rangle-\langle d^{\dagger}_{l\rightarrow}d_{l\rightarrow}\rangle.
   \end{equation}
   \end{subequations}

    It is interesting to note from Eq.\ (\ref{c&s1}) that the ($c$)-fermion charges is not directly proportional to the $d$-fermion occupation number, and can be changed only by exciting a pair of $d$ fermions. This is because the $d$-fermions are equal superposition of particle- and hole- states of $c$-fermions. As a result they carry only spin and no charge individually.

   In terms of $d$'s, $H_0$ becomes
\begin{eqnarray}
\label{href}
H_0 & \rightarrow & 2i\tilde{t}\sum_{\langle i,j\rangle}\left[-(d^{\dagger}_{i1}d^{\dagger}_{j1}-d_{j1}d_{i1})
+\delta(d^{\dagger}_{i2}d^{\dagger}_{j2}-d_{j2}d_{i2})\right]   \\ \nonumber
& = & 2i\tilde{t}\sum_{\langle i,j\rangle}\left[-(d^{\dagger}_{i\leftarrow}d^{\dagger}_{j\rightarrow}-d_{j\rightarrow}d_{i\leftarrow})
+\delta(d^{\dagger}_{i\rightarrow}d^{\dagger}_{j\leftarrow}-d_{j\leftarrow}d_{i\rightarrow})\right]
\end{eqnarray}
 which can be diagonalized straightforwardly by introducing sublattice Fourier transforms
 \[
 d^{A(B)}_{\mathbf{k}\alpha}={1\over V}\sum_{i\in A(B)}e^{i\mathbf{k} \cdot \mathbf{r}_i}d_{i\alpha}  \]
 etc., where $\alpha=1,2$ and
 \begin{subequations}
 \begin{equation}
 \label{hab}
 H_0=\sum_{\mathbf{k},\alpha=1,2}\psi^{\dagger}_{\mathbf{k}\alpha}h^{\alpha}(\mathbf{k})\psi_{\mathbf{k}\alpha},
\end{equation}
 where $\psi^{\dagger}_{\mathbf{k}\alpha}=(d^{\dagger A}_{\mathbf{k}\alpha},d^{B}_{-\mathbf{k}\alpha})$ and
\begin{eqnarray}
h^{\alpha}(\mathbf{k})=\left(
\begin{array}{cc}
0 & \Delta_{\alpha}(\mathbf{k}) \\
\Delta^*_{\alpha}(\mathbf{k}) & 0
\end{array}
\right)
\end{eqnarray}
\end{subequations}
 with $\Delta_1(\mathbf{k})=4it(\sum_{i=1,..dim}\sin k_i)$ where $dim$ is the dimension of the system and $\Delta_2(\mathbf{k})=\delta\Delta_1(\mathbf{k})$.

  Diagonalizing the Hamiltonian, we obtain
  \begin{equation}
  \label{hdia0}
  H_0=\sum_{\mathbf{k},\alpha=1,2}E_{\alpha}(\mathbf{k})\left(\gamma^{\alpha{\dagger}}_{\mathbf{k}+}\gamma^{\alpha}_{\mathbf{k}+}
  +\gamma^{\alpha{\dagger}}_{\mathbf{k}-}\gamma^{\alpha}_{\mathbf{k}-}\right),
  \end{equation}
  where $E_{\alpha}(\mathbf{k})=|\Delta_{\alpha}(\mathbf{k})|$,
  \[
    \gamma^{\alpha}_{\mathbf{k}+}(\gamma^{\alpha{\dagger}}_{-\mathbf{k}-})={1\over\sqrt{2}}(d^{A}_{\mathbf{k}\alpha}-(+) id^{B{\dagger}}_{-\mathbf{k}\alpha}).  \]
   with ground state energy $E_G=-\sum_{\mathbf{k}\alpha}E_{\alpha}(\mathbf{k})$.

We notice that $E_1(\mathbf{k})\neq E_2(\mathbf{k})$, reflecting that the fermion pairing breaks spin-rotation symmetry as can be seen directly from Eq.\ (\ref{href}). The spectrum has a Fermi surface denoted by $E_{1(2)}(\mathbf{k})=0$ which describes a rather unusual {\em gapless} BCS-superconductor\cite{PhysRevB.89.235102}. 

  It is also easy to show that for $\delta\neq0$,
 \begin{equation}
 \label{exp1}
 \langle d^{\dagger}_{l\rightarrow}d_{l\rightarrow}\rangle=\langle d^{\dagger}_{l\leftarrow}d_{l\leftarrow}\rangle= {1\over2}
 \end{equation}
 for $l$ in both sublattices, indicating that the ground state is non-magnetic.

  \emph{Properties of the exact solution (2) - $U\neq0$, $\delta=0$}---In terms of the composite fermions $d$'s, the Hubbard interaction term can be expressed as,
\begin{equation}
\label{interaction}
U\sum_{l}(2i\eta_{l\uparrow}\beta_{l\uparrow})(2i\eta_{l\downarrow}\beta_{l\downarrow}) = U\sum_l(n^{(d)}_{l\rightarrow}-{1\over2})(n^{(d)}_{l\leftarrow}-{1\over2})
\end{equation}
where $n^{(d)}_{l\rightleftarrows}=d^{\dagger}_{l\rightleftarrows}d_{l\rightleftarrows}$. In the limit $\delta=0$, the Hamiltonian in terms of composite fermions become
\begin{equation}
\label{hcf}
H  \rightarrow  -2i\tilde{t}\sum_{\langle i,j\rangle}(d^{\dagger}_{i1}d^{\dagger}_{j1}-d_{j1}d_{i1})+U\sum_{l}((D_l)(n^{(d)}_{l1}-{1\over2})
\end{equation}
 (see Eq.\ (\ref{bdgy2}) for the relation between 1(2) and $\rightleftarrows$) and $D_l=n^{(d)}_{l2}-{1\over2}=\pm{1\over2}$ are conserved quantities which can be determined (in the ground state) by minimizing the energy of the system  (see discussions after Eq.\ (\ref{hma})). We have performed the calculation numerically and find that $D_l$ have uniform value $UD_l=-{|U|\over2}$ on the ground state. In particular, the $U$ and $-U$ ground states are related by flipping $D_l$ to $-D_l$ with the solution for $d_1$ fermions remains unchanged.

 It may be surprising that although the $d$-fermions carry zero $c$-fermion charge, nevertheless they are affected by the presence of Hubbard-interaction as indicated in Eqs.\ (\ref{interaction}) and\ (\ref{hcf}). To clarify this we construct the on-site states with zero, one and two occupied $d$-fermions, respectively. Using Eq.\ (\ref{bdgy}), it is easy to show that
  \begin{eqnarray}
  \label{hibd}
  |0_d\rangle & = & {1\over\sqrt{2}}(1+c^{\dagger}_{\leftarrow}c^{\dagger}_{\rightarrow})|0\rangle,  \\ \nonumber
  |s_{d}\rangle & = &  c^{\dagger}_{s}|0\rangle;   \\ \nonumber
  |\rightleftarrows_d\rangle & = & {1\over\sqrt{2}}(1-c^{\dagger}_{\leftarrow}c^{\dagger}_{\rightarrow})|0\rangle
  \end{eqnarray}
  where $|0\rangle$ denotes vacuum for the $c$-fermions, $|0_d\rangle$ denotes vacuum for the $d$-fermions, $|s_{d} \rangle $ denotes a state occupied by a single $d$ fermion with spin $s=\rightleftarrows$ and $|\rightleftarrows_d\rangle$ denotes a state occupied by two $d$-fermions. Notice that the $c$-fermion number is equal to 1 in all 4 states. However, there exists doubly occupied $c$-fermion state components in states $|0_d\rangle$ and $|\rightleftarrows_d\rangle$, and they are both affected by the Hubbard interaction $U$.

  Fourier transforming, the quasi-particle Hamiltonian\ (\ref{hcf}) in the ground state sector $UD_l=-{|U|\over2}$ can be rewritten as
 $H=\sum_{\mathbf{k}}\psi^{\dagger}_{\mathbf{k}}h^{(1)}(\mathbf{k})\psi_{\mathbf{k}}$, where
 $\psi^{\dagger}_{\mathbf{k}\alpha}=(d^{\dagger A}_{\mathbf{k}1},d^{B}_{-\mathbf{k}1})$, and
\begin{eqnarray}
\label{hint}
h^{(1)}(\mathbf{k})=\left(
\begin{array}{cc}
-{|U|\over2} & \Delta_{1}(\mathbf{k}) \\
\Delta^*_{1}(\mathbf{k}) & {|U|\over2}
\end{array}.
\right)
\end{eqnarray}

The quasi-particle energy spectrum for the $d_1$ fermions is given by
\begin{subequations}
\begin{equation}
  \label{hdia}
  H=\sum_{\mathbf{k}}E_{1}(\mathbf{k})\left(\gamma^{(1){\dagger}}_{\mathbf{k}+}\gamma^{(1)}_{\mathbf{k}+}
  +\gamma^{(1){\dagger}}_{\mathbf{k}-}\gamma^{(1)}_{\mathbf{k}-}\right),
  \end{equation}
  where
\begin{equation}
\label{E1}
E_{1}(\mathbf{k})=\sqrt{|\Delta_1(\mathbf{k})|^2+({U\over2})^2}
\end{equation}
and
\begin{equation}
\label{uv}
\gamma^{(1)}_{\mathbf{k}+}(\gamma^{(1){\dagger}}_{-\mathbf{k}-})=u_{\mathbf{k}+(-)}d^{A}_{\mathbf{k}\leftarrow}-(+) iv_{\mathbf{k}+(-)}d^{B{\dagger}}_{-\mathbf{k}\rightarrow}
\end{equation}
where
\[
u(v)_{\mathbf{k}+}=\sqrt{{1\over2}\left(1-(+){|U|\over 2E_1(\mathbf{k})}\right)}  \]
and $u_{\mathbf{k}-} = v_{\mathbf{k}+}; v_{\mathbf{k}-} = u_{\mathbf{k}+}$. The quasi-particles are chargeless and carry spin $1/2$ along $\hat{y}$-direction.
\end{subequations}

   It is also straightforward to show that
   \begin{eqnarray}
   \label{mag1}
   \langle d^{\dagger}_{l1}d_{l1}\rangle &  = & {1\over2V}\sum_{\mathbf{k}}\left(1+{|U|\over 2E_1(\mathbf{k})}\right)  \\ \nonumber
   \langle d^{\dagger}_{l\rightarrow}d^{\dagger}_{l\leftarrow}\rangle & = & \langle d_{l\leftarrow}d_{l\rightarrow}\rangle = 0
   \end{eqnarray}
   in the ground state for both sublattices $l\in A(B)$ and
   \begin{equation}
   \label{mag2}
   m_y=\frac{1}{2}(\langle d^{\dagger}_{l1}d_{l1}\rangle-\langle d^{\dagger}_{l2}d_{l2}\rangle) ={1\over4V}\sum_{\mathbf{k}}\left({|U|\over 2E_1(\mathbf{k})}+sgn(U)\right)
   \end{equation}
    is the staggered magnetization carried by the ground state (recall that  $d_1=d_{\leftarrow(\rightarrow)}$ in sublattices $A(B)$, respectively). We see that the ground state is spin polarized for any $U\neq0$. The spins are fully polarized in the $U\rightarrow\infty$ limit and the spin polarization approaches zero when $U\rightarrow-\infty$. The singular behavior of magnetization at $U\rightarrow0$ reflects the singular nature of our Hamiltonian in the $\delta\rightarrow0$ limit where all $d_2$ quasi-particles are localized.

    Besides quasi-particle excitations $d_1$, we may create {\em solitonic} excitations by flipping $D_l$'s from ground state. The energy of a single soliton excitation $E_{sol}$ is obtained by calculating the ``ground state" energy of Hamiltonian\ ({\ref{hcf}) with a singly flipped $D_l$. We have performed this calculation numerically in a square lattice for various values of $U/t$ and the results are shown in Fig.(\ref{spin_energy_excitation}). We note that the excitation energy is proportional to $U^2$ at small $U$ but is proportional to $\tilde{t}^2/|U|$ for $|U|\gg\tilde{t}$. Physically, the soliton excitation is created by adding (or subtracting) a localized $d_2$ fermion with a dressed cloud of $d_1$ fermions. The charge and spin carried by the soliton is calculated using Eq.\ (\ref{c&s1}) and we find that the soliton is chargeless and carries spin $1$ for $U\gtrsim4\tilde{t}$ and spin $0$ for $U\lesssim-4\tilde{t}$. There exists also a small region around $|U|\lesssim  4\tilde{t}$ where the spin is $1/2$ (see Fig.(\ref{spin_energy_excitation})). Physically, the soliton is a bound state between the $d_2$ particle and $d_1$ hole when $U/4\tilde{t}$ is large and positive and is a bound state between $d_2$ hole and $d_1$ hole when $U/4\tilde{t}$ is large and negative. The $d_{1(2)}$ fermions are unbounded when $|U|\lesssim4\tilde{t}$. Consequently we expect that the soliton is a boson (spin=$0,1$) when $|U|\gtrsim 4\tilde{t}$ and is a  spin-$1/2$ fermion when $|U|\lesssim 4\tilde{t}$.

      To study the $\delta\rightarrow0$, $U\rightarrow0$ region more carefully we show in Fig.(\ref{spin_energy_excitation}) the ground state staggered magnetization magnitude as a function of $U/\tilde{t}$ in our square lattice model with $\delta=0.0$ and $0.1$ computed using a mean-field approximation
 \begin{eqnarray}
 \label{mft}
  & &(n^{(d)}_{l\rightarrow}-{1\over2})(n^{(d)}_{l\leftarrow}-{1\over2}) \rightarrow  (n^{(d)}_{l\rightarrow}-{1\over2})\langle(n^{(d)}_{l\leftarrow}-{1\over2})\rangle \\ \nonumber
 & & + \langle(n^{(d)}_{l\rightarrow}-{1\over2})\rangle(n^{(d)}_{l\leftarrow}-{1\over2})-\langle(n^{(d)}_{l\rightarrow}-{1\over2})\rangle
 \langle(n^{(d)}_{l\leftarrow}-{1\over2})\rangle \\ \nonumber
 & & = -i \left( c_{l,\uparrow}^{\dagger}c_{l,\downarrow} \mathrm{Im} \left(\langle c_{l,\downarrow}^{\dagger}c_{l,\uparrow} \rangle \right)+\xi_l c_{l,\uparrow}^{\dagger}c_{l,\downarrow}^{\dagger} \mathrm{Im} \left( \langle c_{l,\downarrow}c_{l,\uparrow} \rangle \right) \right) \\ \nonumber
 & &  +h.c.,
 \end{eqnarray}
where we have used Eqs.\ (\ref{fermiony}) and\ (\ref{bdgy}) in deriving the last equality.  The ground state expectation values $\langle(n^{(d)}_{l\rightleftarrows}-{1\over2})\rangle$ are determined self-consistently from the mean field theory. The mean field result becomes exact in the limit $U=0$ and $U/\delta\rightarrow\infty$ and the $\delta\neq0$ mean-field calculation provides an extrapolation between the two exact limits. We find that the singular behavior of staggered magnetization $m_y$ at $\delta=0$ as given by Eq.\ (\ref{mag2}) is smoothed out for $\delta=0.1$. The excitation energies and spins carried by the solitions for $\delta=0.1$ are also calculated in the mean field theory for different values of $U$ and are shown in Fig.(\ref{spin_energy_excitation}) for comparison. We see that both the excitation energies and spins carried by the solition are similar for $\delta=0.0$ and $0.1$.

\begin{figure}[tbh!]
\centering
\mbox{
{\includegraphics[width=0.45\textwidth]{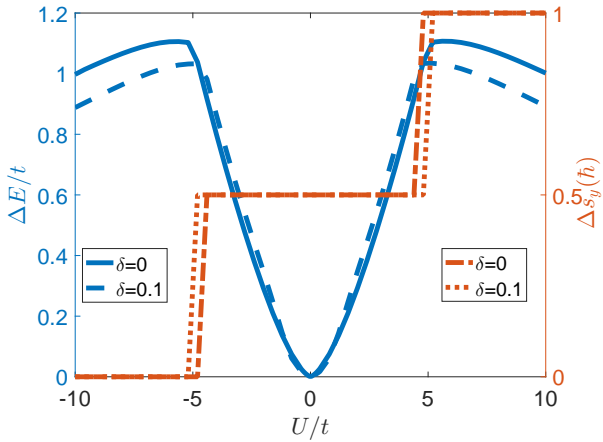}
}
}
\caption{The excitation energy ($\Delta E$ left axis) and spin along $\hat{y}$-direction ($\Delta s_y$ right axis) as a function of $U$ for $\delta=0$ and $0.1$. The blue-solid and blue-dash lines represent the excitation energies for $\delta=0,0.1$, and the orange-dot and orange-dash-dot lines represent the spins carried by the excitations for $\delta=0,0.1$ respectively.}\label{spin_energy_excitation}
\end{figure}

\begin{figure}[tbh!]
\centering
\mbox{
{\includegraphics[width=0.45\textwidth]{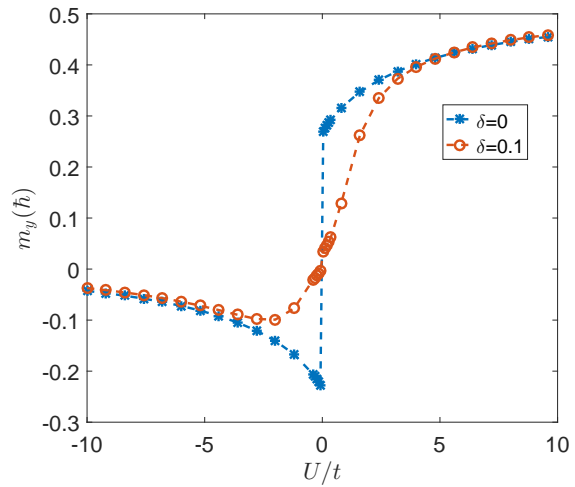}
}
}
\label{magnetization}
\caption{The staggered magnetization in $\hat{y}$-direction $m_y$ as a function of $U$ for different $\delta$'s. The blue-star and orange-circle  lines are  for $\delta=0,0.1$ respectively.}
\end{figure}
\emph{Summary and discussions}---Summarizing, we introduce in this paper an exact solvable BCS-Hubbard model in arbitrary dimensions. The construction of the exact solution is parallel to the exactly solvable Kitaev honeycomb model for $S=1/2$ quantum spins and can be viewed as a generalization of Kitaev's construction to $S=1/2$ interacting lattice fermions. In fact, any Hamiltonian which when represented in terms of Majorana fermions, has the form
\begin{equation}
\label{more}
H=4i\sum_{\langle i,j\rangle,\sigma}\left(t_{ij}\beta_{i\sigma}\beta_{j\sigma}\right)+U\sum_{l}(2i\eta_{l\uparrow}\beta_{l\uparrow})(2i\eta_{l\downarrow}\beta_{l\downarrow}),
\end{equation}
 is exactly solvable following our discussion on the construction on exact solution, independent of dimension and lattice structure. The nearest neighbor hopping ($t$) + pairing ($\Delta$) model on square (and cubic) lattices we consider in this paper is just an example of a large class of exactly solvable lattice fermion models that can be written in the form\ ({\ref{more}). We also notice that the fermionic form Eq.(\ref{hcf}) is similar to the Falicov-Kimball model\cite{PhysRevLett.22.997,prosko2017simple}.

 Physically, the presence of the ESP pairing term $\Delta$ breaks spin rotation symmetry making one of the two quasi-particle bands completely flat in the limit $\delta\rightarrow0$. The quasi-particles in the flat band are localized making the resulting Hamiltonian exactly solvable. The same happens in the Kitaev honeycomb model. We note that the quasi-particles are non-perturbative objects that are related to the original $c$-fermion states by a {\em local} Bogoliubov-de Gennes transformation\ (\ref{bdgy}) in our model. As a result the quasi-particle and solitonic  excitations both carry non-trivial charge and spin quantum numbers as discussed in the main text.

 Lastly we comment that our construction of exactly solvable model suggests a new mean field decoupling channel of Hubbard interaction\ (\ref{mft}) which can be applied to any interacting fermion model when expressed in terms of Majoranan fermions. The decoupling scheme breaks spin-rotation symmetry and becomes exact when one of the quasi-particle band becomes flat.

\emph{Acknowledgment}---Z. Chen and X. Li acknowledge Dr. Yao Lu for helpful discussions. This work is supported by Hong Kong RGC through grant HKUST3/CRF/13G.
\bibliographystyle{apsrev4-1}
\bibliography{ref_BCS_Hubbard}
\end{document}